\documentclass[11pt,preprint]{aastex}

\shorttitle{20-100 keV excess in NGC 1365}
\shortauthors{G. Risaliti et al.}

\begin{document}

\title{A strong excess in the 20-100 keV emission of NGC 1365}

\author{G. Risaliti\altaffilmark{1,2},
V. Braito\altaffilmark{3,4},
V. Laparola\altaffilmark{5},
S. Bianchi\altaffilmark{6},
M. Elvis\altaffilmark{1}, 
G. Fabbiano\altaffilmark{1}, 
R. Maiolino\altaffilmark{7},
G. Matt\altaffilmark{6},
J. Reeves\altaffilmark{8},
M. Salvati\altaffilmark{2},
J. Wang\altaffilmark{1}
}
\email{grisaliti@cfa.harvard.edu}

\altaffiltext{1}{Harvard-Smithsonian Center for Astrophysics, 60 Garden St. 
Cambridge, MA 02138 USA}
\altaffiltext{2}{INAF - Osservatorio di Arcetri, Largo E. Fermi 5,
I-50125, Firenze, Italy}
\altaffiltext{3}{Department of Physics and Astronomy, University of Leicester, University Road, Leicester, LE1 7RH, UK}
\altaffiltext{4}{Department of Physics and Astronomy, Johns Hopkins University, Baltimore, MD 21218}
\altaffiltext{5}{INAF - -IASF, Via U. La Malfa 153, 90146 Palermo, Italy}
\altaffiltext{6}{Dipartimento di Fisica, Universit\`a degli Studi ``Roma Tre'',
Via della Vasca Navale 84, I-00146 Roma, Italy}
\altaffiltext{7}{INAF -
Osservatorio Astronomico di Roma
via di Frascati 33
00040 Monte Porzio Catone, Roma, Italy}
\altaffiltext{8}{Astrophysics Group, School of Physical and Geographical Science, Keele University, Keele, Staffordshire ST5 5BG, UK}

\begin{abstract}
We present a new {\em Suzaku} observation of the obscured AGN in NGC~1365, revealing an
unexpected excess of X-rays above 20~keV of at least
a factor $\sim$2 with respect to the extrapolation of the best-fitting 3-10~keV model. 
Additional {\em Swift-BAT} and
{\em Integral-IBIS} observations show that the 20-100~keV is concentrated within $\sim$1.5~arcmin
from the center of the galaxy, and is not significantly variable on time scales from days to
years. A comparison of this component
with the 3-10~keV emission, which is characterized by a rapidly variable absorption, suggests
a complex structure of the circumnuclear medium, consisting of at least two distinct
components with rather different physical properties, one of which covering $>$80\% of the
source with a column density $N_H$$\sim$3-4$\times$10$^{24}$~cm$^{-2}$. An alternative explanation is the
presence of a double active nucleus in the center of NGC~1365.
\end{abstract}


\section{Introduction}

X-ray spectroscopy is a powerful way to investigate the complex structure of
the circumnuclear medium in Active Galactic Nuclei. 

In particular, the energy window above 10~keV is crucial to understand
the nature of Compton-thick AGNs, i.e. those absorbed by a 
column density higher than 10$^{24}$~cm$^{-2}$. 
These objects 
represent about 40-50\% of all AGNs
in the local Universe  (Risaliti et al.~1999, Guainazzi et al.~2005) and 
are expected to be also common at higher redshifts (Gilli et al.~2007).
If the absorbing column density is in the range $10^{24}$$<$N$_H$$<$10$^{25}$~cm$^{-2}$, 
%
%
the primary AGN emission
becomes directly observable only above 10~keV, while only the much fainter reflected 
emission is visible below 10~keV. 
This is the case, for example, of two out of the three nearest AGNs, NGC 4945
(Guainazzi et al. 2000) and Circinus (Matt et al.~1999). 


Recently, several new instruments increased our ability to perform such studies. In particular,
{\em Suzaku} is equipped with the most sensitive instrument available in the 15-60~keV interval,
while {\em Integral}-IBIS and {\em Swift}-BAT are performing all-sky surveys in the 15-150~keV
band, providing good quality time-averaged spectra and positional accuracy of the order of few arcmin.

In this Letter we present the discovery of a new case of a strong excess in the $>$10~keV emission of
an AGN, characterized by a so-far unique, puzzling property: the source showing this excess, NGC~1365,
is not reflection-dominated in the 2-10~keV range. Instead, it is a well known Compton-thin source,
intensively studied by our group for its extreme absorption variability, due to  
clouds with column density from a few 10$^{23}$ to $>$10$^{24}$~cm$^{-2}$ covering and uncovering the
X-ray source on timescales as short as $\sim$10~hours (Risaliti et al.~2009A, 2009B).

\section{The {\em Suzaku} Observation}

{\em Suzaku} observed NGC~1365 (in the XIS nominal position) on January 21-25, 2007 for
an elapsed time of $\sim$3~days, and a net time of 150~ks.
 
{\em Data Reduction.} 
The spectra and calibrations 
were obtained following the standard
procedure illustrated in the {\em Suzaku} reduction guide\footnote{http://heasarc.gsfc.nasa.gov/docs/suzaku/analysis/abc/}, and using the most recent calibrations.

For the low-energy instruments (XIS) the source spectrum was extracted from
a circular region with radius of 2.9~arcmin, centered on the source. The background
was obtained from a free region in the same field of view. 
Calibration files were produced using the FTOOLS 6.6 package\footnote{http://heasarc.gsfc.nasa.gov/docs/software/ftools/ftools\_menu.html}. The spectra and calibrations from the two front-illuminated CCDs 
(XIS0 and XIS3) were merged, while the spectrum from the back-illuminated XIS1 detector 
has been treated 
separately in the subsequent spectral analysis. 
The soft (0.5-2.5~keV) emission of NGC~1365 is know to be dominated by a thermal component
originating from the diffuse gas in the center of the galaxy (Wang et al.~2009). 

For the HXD-PIN data,  
we used the rev2 data, which include all 4 cluster units, and the
best available background (Fukazawa et al.~2009), which accounts for  the instrumental background
(NXB, Takahashi et al.~2007, Kokubun et al.~2007), and is affected by systematic uncertainties of
about 1.3\% (at 1~$\sigma$). We  then
simulated a spectrum for the cosmic X-ray background counts (Boldt~1987, Gruber et al.~1999) and
added  it to the  instrumental one.
Using this background NGC1365 is detected in the 15--70 keV band 
 at $\sim 20$\%  above the
background  with a net count rate  of  (8.7$\pm$0.2)$\times$10$^{-2}$ 
(a total of  $\sim$12000 net source counts have been collected).

The total 3-10~keV and 15-60~keV spectra are shown in Fig.~1.
In the spectral analysis, we used a cross-calibration constant of 1.16 between
the HXD and XIS spectra, as suggested by the {\em Suzaku}-HXD calibration team\footnote{
http://www.astro.isas.jaxa.jp/suzaku/doc/suzakumemo/suzakumemo-2008-06.pdf}.
\\

\noindent {\em Data analysis.} 
A visual inspection of the low-energy spectrum suggests that the source is in a Compton-thin state,
with a photoelectric cut-off at 3-4~keV, corresponding to an 
absorbing column density of the order of 1-2$\times10^{23}$~cm$^{-2}$. The spectral shape, and the 
value of the total flux, closely resembles the state observed during the {\em XMM-Newton} observation 
described in Risaliti et al.~2009B. We therefore used the same model developed to fit those data
as a starting point for our spectral analysis. The model consists of the following
components: a continuum, made by a power law absorbed by a double layer of gas, one layer completely covering the source, 
and one layer covering only a fraction of it; a cold reflection continuum, a relativistically broadened
iron emission line, and a set of four absorption lines in the 6.7-8.3~keV range, due to
Fe~XXV and FeXXVI K$\alpha$ and K$\beta$ transitions. {\bf A complete fit of the whole XIS band (0.5-10~keV)
would also include a soft thermal component, dominant in the soft band (0.5-2~keV). This component
has a diffuse origin, as shown by our high-resolution {\em Chandra} observations (Wang et al.~2009), and
its contribution at E$>$3~keV is completely negligible, so we did not include it in our model.}

This model provided a good fit to the low-energy data ($\chi^2$=1215/1182~d.o.f.) with no obvious residual features
(Fig.~1). All the components listed above are statistically significant, with no redundancy.
However, when compared with the high-energy spectrum, we find that the extrapolation of the
3-10~keV best fit model falls short  by a factor of $\sim$3 in reproducing the 15-60~keV emission (Fig.~1).

A complete analysis of the 3-10~keV emission, focusing on the complex (and variable, see below) 
absorbing medium will be presented
in a forthcoming paper (Maiolino et al.~2009, in prep.). Here instead we focus on the 
comparison between the low- and high- energy time-averaged spectra. As a consequence, our spectral analysis 
of the XIS spectrum is mainly aimed at determining the continuum parameters, in order to compare the
high energy spectrum with the extrapolation of the emission below 10~keV.

We checked several possibilities in order to explain the discrepancy shown in Fig.~1:\\
\underline{1) Continuum slope and flux.} We fitted the low- and high- energy simultaneously, in order
to search for an acceptable fit of both components. Qualitatively, we may expect that a
much flatter continuum, with a higher extrapolation at $>$15~keV, could reproduce the
high energy emission. No satisfactory solution can be found in this scenario: {\bf it is impossible
to obtain a good fit of the average continuum below 10~keV with any slope $\Gamma<2.2$. 

However, given the strong spectral variability below 10~keV, this result may change when the 
different spectral states are fitted separately. 
In particular, the work performed on the Compton-thin spectrum observed in January 2004 by {\em XMM-Newton}
(Risaliti et al.~2009B) shows that an accurate, time-resolved analysis is necessary in order
to correctly estimate the relevant parameters of several spectral components, such as the
broad iron emission line, the reflection component, and the structure of the absorber.
The {\em Suzaku} observation is no exception: a time resolved analysis reveals 
 significant absorption variability.
This issue is discussed in detail in Maiolino
et al.~2009 (in prep.), where it is shown that a flatter photon index ($\Gamma\sim2$) can indeed
reproduce the continuum if the absorption component is free to vary during the observation. However,
even in this case, a strong high-energy excess is still present, with an observed 15-60~keV flux 
larger than the extrapolated lowe-energy emission by a factor of two.}   

\underline{2) Additional continuum component.} A good fit of the high energy excess can be obtained adding a
 continuum component absorbed by a column density $N_H$$\gtrsim$10$^{24}$~cm$^{-2}$.
We modeled this emission with a power law with the same spectral slope as the 3-10~keV primary component and,
as a first step, with just photoelectric absorption. The best fit value for the column density is 
$N_H$=3.8$\pm0.5$$\times$10$^{24}$~cm$^{-2}$, while the flux of the extra component is a factor $\sim$3
higher than that of the low-energy component. All the best fit parameters of the other components
remain the same as in the fit of the low-energy component only. This is also the case for the
other models of high-energy absorption discussed below. Therefore, the parameters listed in the upper part of
Table~1, regarding the low-energy components, are common for all the models discussed here.

A column density $N_H$$\sim$4$\times$10$^{24}$~cm$^{-2}$ corresponds to a Compton optical depth
$\tau_C$$\sim$2.5. Therefore, the effects of Compton scattering must be taken into account in order
to estimate the correct intrinsic flux and column density. In particular, the intrinsic flux
obtained in the previous fit, with the photoelectric absorption only, can be regarded as a lower
limit of the actual emission of the high-energy component. We repeated the previous fit adding a Compton
attenuation factor exp(-$\tau_C$). Physically, this represents a scenario where every Compton-scattered photon 
is removed from the observed beam. This happens if the $N_H$=3-4$\times$10$^{24}$~cm$^{-2}$ gas is
{\em only} along the line of sight, and covers a negligible fraction of the solid angle as viewed from the 
X-ray source. The results of the fit with  this model are a huge intrinsic flux (about 30 times more than the
low-energy component) and a column density consistent with the previous value. This shows that the 
determination of the flux level is critically dependent on the geometry of the absorber/scatterer, while
the column density value is mainly constrained by the spectral curvature around 15-30~keV, and
is therefore well determined independently from the assumed geometry.
A more physical treatment of Compton scattering has been made by Matt, Pompilio \& La~Franca~(1999) and 
Yaqoob et al.~(1997), assuming a 
spherically symmetric, homogeneous gas around an X-ray source with a power law emission. The latter model
is implemented in XSPEC, and provides a correct output in the range 12-19~keV. We therefore repeated
our analysis using only the 14-19~keV high energy data, and freezing the column density to the
value obtained in the previous fits. The result is an intermediate value of the intrinsic flux (about 7 times the
low-energy component, Tab.~1). This, again can be regarded as a lower limit, since a smaller covering factor
of the circumnuclear gas would imply a lower contribution from photons scattered to our line of sight.
 
\begin{table}
\centerline{\begin{tabular}{lc}
$\Gamma^a$ &  2.34$^{+0.03}_{-0.02}$ \\
F$^b_{2-10}$(OBS) & 1.3$\times10^{-11}$ erg s$^{-1}$ cm$^{-2}$ \\
F$^c_{2-10}$(INTR) & 3.9$\times10^{-11}$ erg s$^{-1}$ cm$^{-2}$ \\
F$^d_{20-100}$(EXTR) & 2.2$\times10^{-11}$ erg s$^{-1}$ cm$^{-2}$ \\
L$^d_{20-100}$(EXTR) & 8.7$\times10^{41}$ erg s$^{-1}$  \\
$\chi^2$/d.o.f.  & 512/475\\
\hline
$N_H^e$(high) & 4.5$^{+1.5}_{-0.7}\times10^{24}$ cm$^{-2}$\\
F$^b_{20-100}$(OBS) & 7.2$\times10^{-11}$ erg s$^{-1}$ cm$^{-2}$ \\
F$^{f}_{20-100}$(INTR,$\Omega$=0) & 70$\times10^{-11}$ erg s$^{-1}$ cm$^{-2}$ \\  
L$^{f}_{20-100}$(INTR,$\Omega$=0) & 3.0$\times10^{43}$ erg s$^{-1}$  \\  
F$^{g}_{20-100}$(INTR,$\Omega$=4$\pi$) & 14$\times10^{-11}$ erg s$^{-1}$ cm$^{-2}$\\
L$^{g}_{20-100}$(INTR,$\Omega$=4$\pi$) & 6.1$\times10^{42}$ erg s$^{-1}$ \\
\hline  
\end{tabular}}
\caption{\footnotesize{Main spectral parameters from the best fit of the {\em Suzaku} observation
of NGC~1365. Notes: $^a$: photon index of the low-energy continuum and the high-energy extra component. 
$^b$: observed flux; $^c$: intrinsic (de-absorbed) flux; $^d$: extrapolated 20-100~keV flux and luminosity from the 
low-energy best fit model;
$^e$: column density absorbing the high-energy component; $^f$: intrinsic flux and luminosity 
assuming a negligible coverage of
the 
gas obscuring the high-energy component, i.e. a single cloud along the line of sight; $^g$: same, assuming 
full coverage of the obscuring gas. }}
\end{table}

\section{BeppoSAX, Integral and Swift Observations}

NGC~1365 has been observed in the hard ($>$10~keV) range by other observatories, providing important
complementary information. 

{\bf BeppoSAX}. The first high-energy observation was performed by BeppoSAX in 1999. The 20-100~keV data 
were presented in Risaliti et al.~2000, but not discussed in detail, for two main reasons: the low
S/N of the spectrum, which made the excess barely significant, and the large field of view of
the PDS instrument (about 55 arcmin FWHM), which implies a non-negligible probability of a contribution
of serendipitous sources (in this case, for example, the AGN in NGC~1386, at a distance of about 30 arcmin).
However, the observed emission is roughly compatible with that observed with {\em Suzaku}, with an
average flux $\sim$15\% higher, but compatible within the {\em BeppoSAX}-PDS calibration uncertainties (Fig.~3).

{\bf Integral}. 
Up to now, the sky region of NGC~1365 has been little observed by {\em Integral}. For this reason,
NGC~1365 is not in the catalog of {\em Integral} bright sources (Bird et al.~2007). However, it has
been detected with a signal-to-noise S/N$\sim$5 in the 15-195~keV range by Sazonov et al.~2007, 
with a flux level perfectly
compatible with that measured by {\em Suzaku}, assuming the same spectral slope as obtained in our fits.  
The positional accuracy of the {\em Integral} observation is about 2.5~arcmin, and the source position is compatible
with the nucleus of NGC~1365 (Fig.~2).

{\bf Swift}. The {\em Swift}-BAT instrument observes about 1/6 of the sky continuously, in an almost random way.
The 39-months Palermo-BAT catalog (Cusumano et al.~2009),
provides a spectrum of NGC~1365 with S/N=28, and a positional accuracy of 1.5~arcmin (Fig.~3), compatible  
 with the center of the galaxy. 
The 
average spectrum 
is shown in Fig.~3, and is compatible with the ones obtained with {\em BeppoSAX} and {\em Suzaku} 
in single observations. 
This suggests the absence of significant intrinsic variability in the high-energy emission. Moreover,
by constraining the position of the high energy emission within 1.5 from the nucleus, the {\em Swift}-BAT
data imply that
contamination from hard X-ray bright sources is unlikely also for the Suzaku PIN data.
 
\begin{figure}
\includegraphics[width=13.5cm,angle=0]{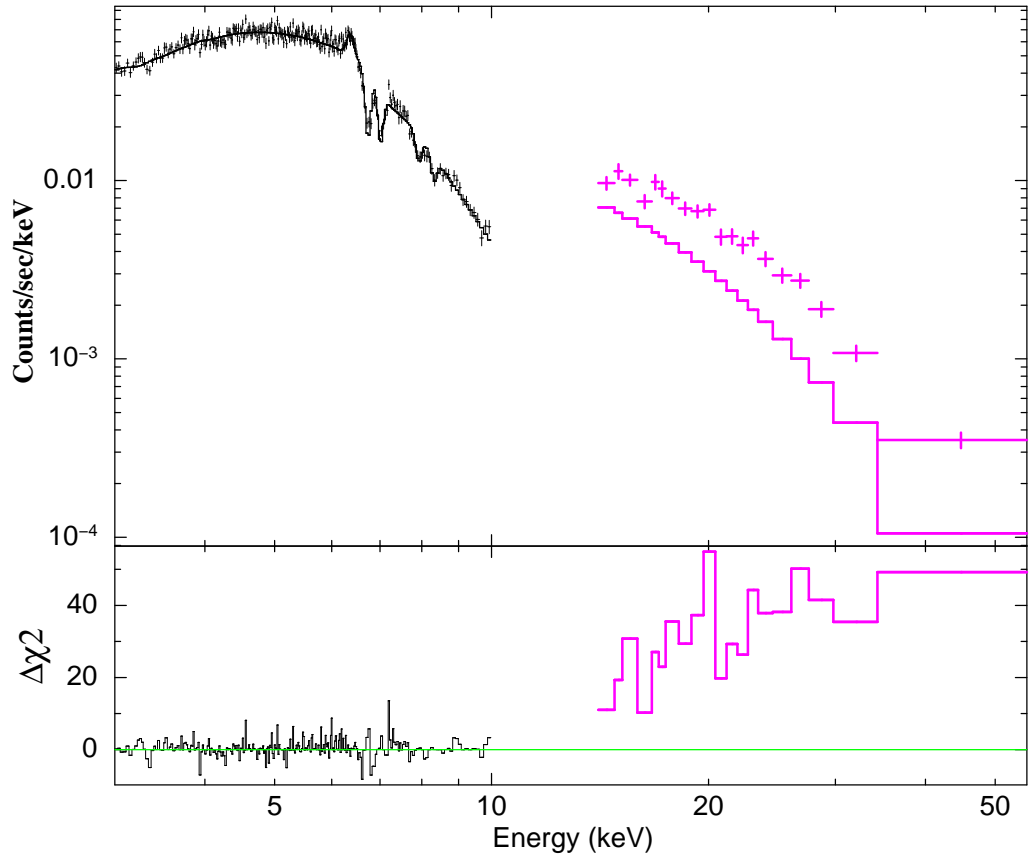}
\figcaption{\footnotesize{
Upper panel: {\em Suzaku} spectrum of NGC~1365, 
and model obtained from the fit of the 3-10~keV component only. Lower panel: residuals. For the
low-energy spectrum, we only show the merged XIS0+XIS3 data.
}}
\label{totfit}
\end{figure}

\section{Discussion}

The hard X-ray analysis of NGC~1365 shows a strong excess in the emission above 10~keV with respect to
the extrapolated lower energy spectrum. The comparison between single {\em Suzaku} and {\em BeppoSAX} observations,
and average {\em Integral} and {\em Swift} ones shows that this emission remains about constant (both in flux and shape, i.e. $N_H$) on time scales of a few years. The {\em Swift}-BAT observations constrain the high-energy source position within 1.5~arcmin from the 
nucleus.

The presence of the extra high-energy component in NGC~1365 is unique among known Compton-thin AGNs, and its interpretation
is quite challenging. 

Several AGNs are much brighter above 10~keV than expected from the extrapolation of their emission at lower energies.
However, in all known cases this is due to a complete coverage of gas with column density in the range 10$^{24}$-10$^{25}$~cm$^{-2}$, resulting in a reflection-dominated spectrum below 10~keV, and an intrinsic component showing up above 10~keV.
NGC~1365 is instead clearly Compton-thin below 10~keV in the {\em Suzaku} observation. The most obvious explanation of the observed emission is therefore a partial covering by a $N_H$$\sim$4$\times$10$^{24}$~cm$^{-2}$ gas,
with a fraction of this intrinsic emission not covered, and absorbed only by clouds with lower column density.

However, this simple scenario is challenged by the X-ray observational history of NGC~1365. The main two relevant facts are the
following:\\
1. NGC~1365 has been observed about 15 times in the 2-10~keV range, with all the main X-ray observatories.
The source was in a Compton-thin state in about 2/3 of the observations, and in Compton-thick, reflection dominated state
in the remaining 1/3. Column density variations, due to both Compton-thin and Compton-thick clouds have been observed
on time scales as short as a few hours. The {\em Suzaku} observation presented here is no exception, with 
$N_H$ changes of a few 10$^{23}$~cm$^{-2}$ in a few hours (Maiolino et al.~2009, in prep). \\
2. In all the observations in a Compton-thin state, the measured intrinsic 2-10~keV flux is constant within a factor $\sim2$,
between $\sim$2 and 4$\times10^{-11}$ erg s$^{-1}$ cm$^{-2}$.
In this range, the {\em Suzaku} observation caught the source at about its recorded maximum.

These data are impossible to reconcile with a single absorber made of clouds with different column densities:
considering that the intrinsic high-energy flux is at least a few times larger than the one at low energy, we
should have an absorber with $N_H$$>10^{24}$~cm$^{-2}$ covering $>$85\% of the source with very small variations in the
covering factor, and at the same time producing the huge and rapid $N_H$ changes observed in the uncovered part.

We are currently developing simulations to reproduce the observed $N_H$ pattern observed in NGC~1365 through a
clumpy absorber, allowing
for different column density distributions for the absorbing clouds, different cloud sizes and variable number of clouds along the line of sight. Given the large number of free parameters, the results are still not unique (we expect to obtain
tighter constraints from a forthcoming survey of 20 snapshot  observations planned with {\em Swift} during 2009). 
However, it is clear that no single   
distribution of clouds can reproduce, through random variations, an almost constant covering of more than 85\% of the source and strong and frequent variations in the remaining $<$15\%.

These considerations imply that only more complex, fine-tuned scenarios are possible:\\
\underline{1. A second, independent absorber.} The constant partial covering could be obtained through an absorber with
physical properties completely different from those of the one responsible for the $N_H$ variations.
A homogeneous screen external to the broad line clouds (most likely responsible for the
variable absorption, Risaliti et al.~2009B) requires an extreme fine tuning on the position with respect
to the observer: the linear dimensions of the X-ray source $D_S$ are of the order of a few 10$^{13}$~cm, while
the broad line region at a distance from the center of the order of R=10$^{16}$~cm. The external 
absorber should therefore have a sharp edge, and 
cover exactly 85-90\% of the source, being at a distance at least $\sim$1000 times
larger than the source size. An alternative scenario is that of an absorber made of many small and dense clouds.
Each cloud should have a linear size $D_C<<D_S$,  and the average number $N$ of clouds along the line of
sight should  satisfy N$\times$($D_S$/$D_C$)$^2$$\sim$0.85-0.9 (in order to obtain the requested covering factor) and
1/$\sqrt{N}$$<$5\%, i.e. N$>$400 (in order to have small fluctuations). Since $D_S$$<$10$^{14}$~cm, the
density of the single clouds should be n$\sim$$N_H$/$D_C$$>10^{12}$~cm$^{-3}$.\\
\underline{2. A warped disk.} The above problems on the fine-tuning of the properties of a homogeneous absorber could be
less severe with an absorber located much nearer to the X-ray source. A warped disk, with one edge crossing
the line of sight to the X-ray source would be a geometrically acceptable solution. 
With a column density
of 4$\times$10$^{24}$~cm$^{-2}$, in order to have a complete absorption below 10~keV,
a ionization parameter $U_X$$<$10$^3$ is needed. With a 
ionizing luminosity L(1-100~keV)$\sim$10$^{43}$~erg~s$^{-1}$ and 
a distance of the order of 10$^{15}$~cm, 
this implies a density n$>$10$^{10}$~cm$^{-3}$, and a disk thickness $\Delta$R/R$\sim$0.4.
This is a plausible geometry, which however requires a fine-tuned relative inclination
between the disk and the line of sight.
\\
\underline{3. A double nucleus.} A second, hidden X-ray source in the nucleus of NGC~1365 is in principle a possible explanation.
Given the high luminosity of the obscured source (L(2-10)$\sim$2$\times$10$^{43}$~erg~s$^{-1}$) this would
be a second AGN. NGC~1365 is a prototype barred spiral, with no signs of recent mergers. This would be the first 
case of a double AGN in a galaxy with a regular morphology. {\em Chandra} and HST observations (Wang et al.~2009) 
do not reveal any indication of such a double nucleus in the X-rays and near-IR, 
down to scales of $\sim$0.5~arcsec, corresponding
to 100~pc. For comparison, we note that NGC 6240, hosting a well known double AGN, is a merging system, with a
distance between the two nuclei of the order of $\sim$1~kpc (Komossa et al.~2003, Risaliti et al.~2006).\\
\underline{4. A serendipitous quasar.} The uniqueness of the case under investigation prompted us to check the
probability of a serendipitous obscured quasar along the line of sight. Assuming the most recent X-ray
luminosity functions and estimates of the density of obscured quasars (Gilli et al.~2007) we obtain that
the probability of a serendipitous quasar with a 20-100~keV flux equal to, or higher than the one of the extra
component at high energy, at a distance higher than that of NGC~1365, in a circular
region with a radius 1.5~arcmin, is about 1.5$\times$10$^{-7}$. 

Our conclusion is that the most likely scenario 
for a single X-ray source is that of two distinct absorbers, 
one made of broad line clouds with $N_H$$\sim$10$^{23}$-10$^{24}$~cm$^{-2}$, sizes of
the order of 10$^{13}$-10$^{14}$~cm and densities n$\sim$10$^{11}$-10$^{12}$~cm$^{-3}$, and a second one 
with $N_H$$\sim$4$\times$10$^{24}$~cm$^{-2}$, consisting either of the outer region of a warped accretion disk, or
of a large number of small, dense clouds with linear dimensions $D_C$$<$10$^{12}$~cm  and densities n$>$10$^{12}$~cm$^{-3}$.

\begin{figure}
\epsscale{0.6}
\includegraphics[width=13.5cm,angle=0]{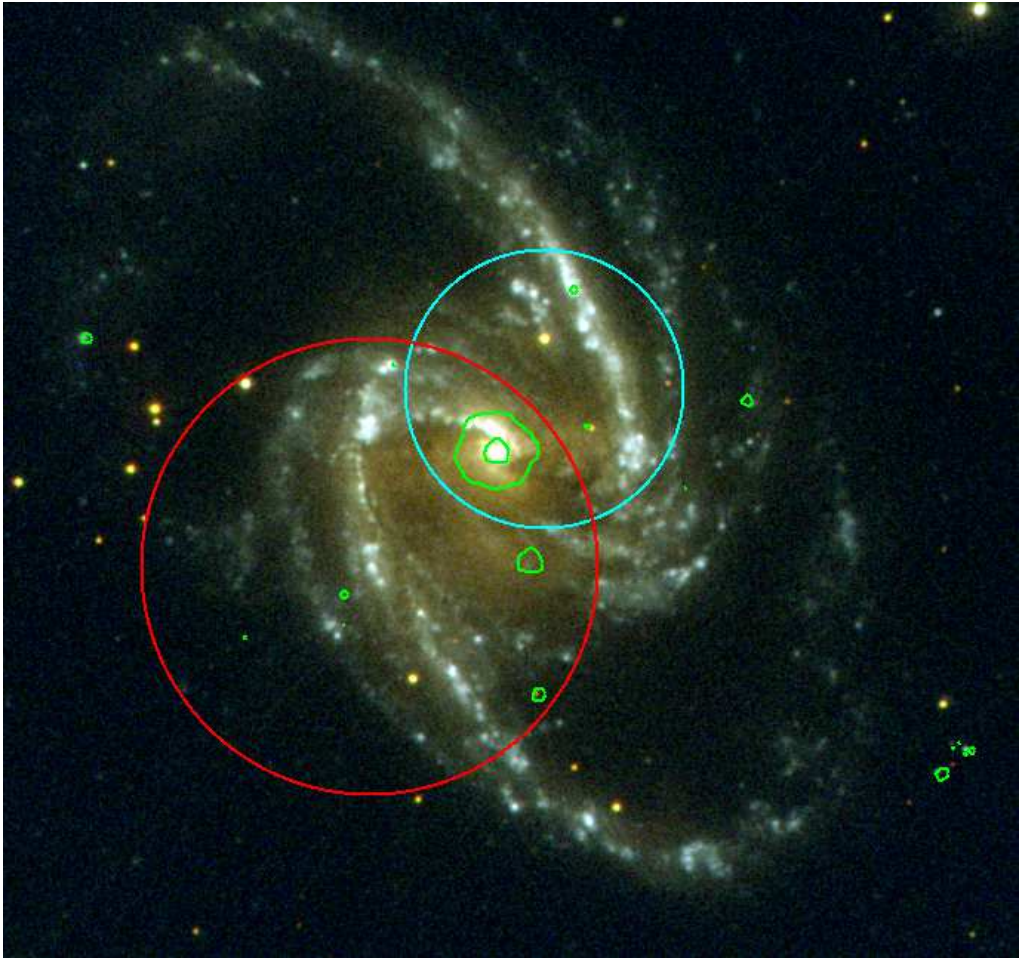}
\figcaption{\footnotesize{
Ultraviolet image of NGC~1365 obtained with the {\em XMM-Newton} optical monitor during the
370~ks observation of May~2008. The green contours show the X-ray sources in the field, as
observed by the EPIC instrument in the same observation. The two circular regions show the 
high energy emission regions obtained from hard X-ray
instruments:
{\em Integral}-IBIS (larger, red circle, Sazonov et al.~2007), and {\em Swift}-BAT
(smaller,  cyan circle, Cusumano et al.~2009).
}}
\label{f2}
\end{figure}

\begin{figure}
\epsscale{0.6}
\includegraphics[width=14.0cm,angle=0]{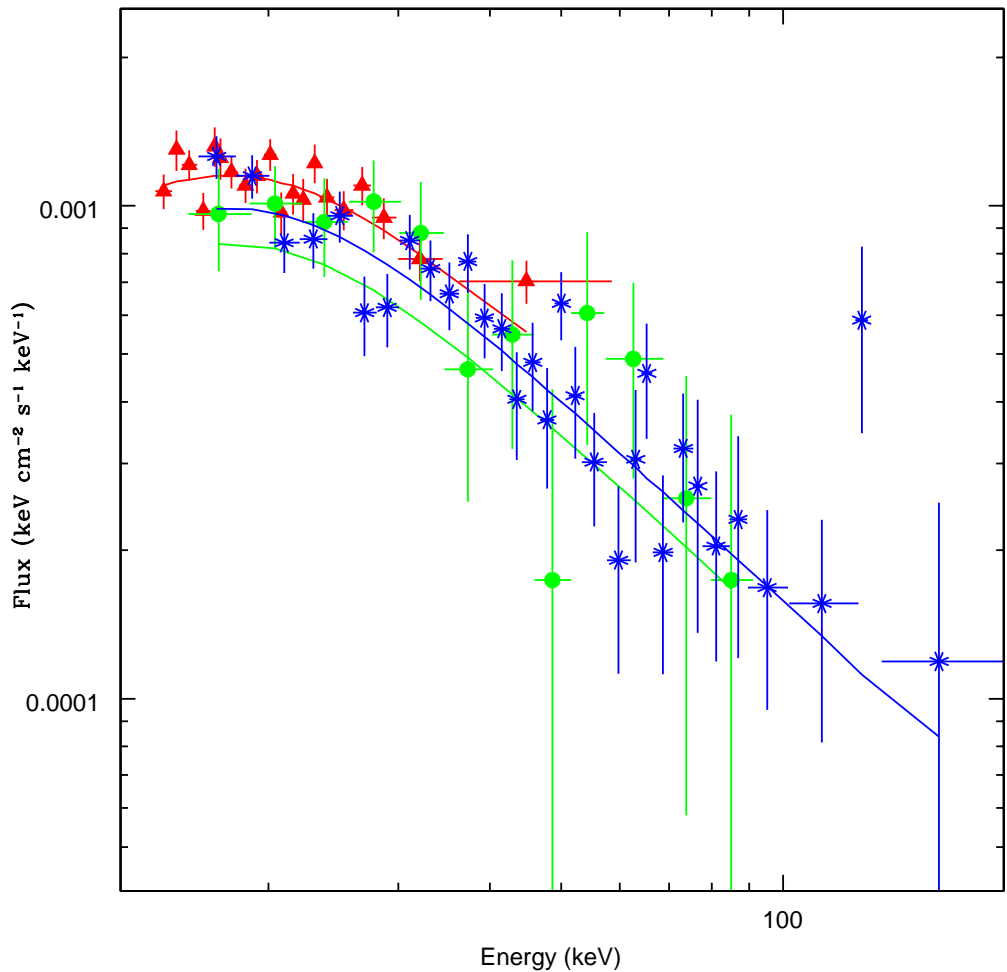}
\figcaption{\footnotesize{Hard X-ray emission of NGC~1365 from {\em BeppoSAX} 
(green circles), {\em Suzaku} (red triangles) and {\em Swift}-BAT (blue stars).
The best fit model, including the high-energy absorbed power law, is also shown.
The three different levels are due to the slightly different calibration
constants 
with respect to the low-energy emission (0.85 for {\em BeppoSAX}, 1.0 for {\em Swift}-BAT, 1.16 for
{\em Suzaku}).
}}
\label{f3}
\end{figure}
 
\acknowledgements

This work has been partly 
supported by grants prin-miur 2006025203, ASI-INAF I/088/06/0, and NASA NNX08AN48G.




\begin{thebibliography}{}
\bibitem[Arnaud(1996)]{1996ASPC..101...17A} Arnaud, K.~A.\ 1996, 
Astronomical Data Analysis Software and Systems V, 101, 17
\bibitem[Bird et al.(2007)]{2007ApJS..170..175B} Bird, A.~J., et al.\ 2007, 
\apjs, 170, 175
\bibitem[Boldt(1987)]{1987PhR...146..215B} Boldt, E.\ 1987, \physrep, 146, 
215
\bibitem[Cusumano et al.(2007)]{} Cusumano, G., et al. 2009, arXiv:0906.4788 
\bibitem[Fukazawa et al.(2009)]{2009PASJ...61S..17F} Fukazawa, Y., et al.\ 
2009, \pasj, 61, 17
\bibitem[Gilli et 
al.(2007)]{2007A&A...463...79G} Gilli, R., Comastri, A., \& Hasinger, G.\ 2007, \aap, 463, 79 
\bibitem[Gruber et al.(1999)]{1999ApJ...520..124G} Gruber, D.~E., Matteson, 
J.~L., Peterson, L.~E., \& Jung, G.~V.\ 1999, \apj, 520, 124
\bibitem[Guainazzi et 
al.(2000)]{2000A&A...356..463G} Guainazzi, M., Matt, G., Brandt, W.~N., Antonelli, L.~A., Barr, P., \& Bassani, L.\ 2000, \aap, 356, 463 
\bibitem[Guainazzi et 
al.(2005)]{2005A&A...444..119G} Guainazzi, M., Matt, G., \& Perola, G.~C.\ 2005, \aap, 444, 119
\bibitem[Kokubun et al.(2007)]{2007PASJ...59S..53K} Kokubun, M., et al.\ 
2007, \pasj, 59, 53
\bibitem[Matt et al.(1999)]{1999NewA....4..191M} Matt, G., Pompilio, F., 
\& La Franca, F.\ 1999, New Astronomy, 4, 191
\bibitem[Matt et 
al.(1999)]{1999A&A...341L..39M} Matt, G., et al.\ 1999, \aap, 341, L39 
\bibitem[Risaliti et al.(1999)]{1999ApJ...522..157R} Risaliti, G., 
Maiolino, R., \& Salvati, M.\ 1999, \apj, 522, 157
\bibitem[Risaliti et 
al.(2000)]{2000A&A...356...33R} Risaliti, G., Maiolino, R., \& Bassani, L.\ 2000, \aap, 356, 33 
\bibitem[Risaliti et al.(2009)]{2009ApJ...696..160R} Risaliti, G., et al.\ 
2009A, \apj, 696, 160
\bibitem[Risaliti et al.(2009)]{2009MNRAS.393L...1R} Risaliti, G., et al.\ 
2009B, \mnras, 393, L1
\bibitem[Wang et al.(2009)]{2009arXiv0901.0297W} Wang, J., Fabbiano, G., 
Elvis, M., Risaliti, G., Mazzarella, J.~M., Howell, J.~H., 
\bibitem[Sazonov et 
al.(2007)]{2007A&A...462...57S} Sazonov, S., Revnivtsev, M., Krivonos, R., Churazov, E., \& Sunyaev, R.\ 2007, \aap, 462, 57 
\bibitem[Takahashi et al.(2007)]{2007PASJ...59S..35T} Takahashi, T., et 
al.\ 2007, \pasj, 59, 35 
\bibitem[Wang et al.(2009)]{2009ApJ...694..718W} Wang, J., Fabbiano, G., 
Elvis, M., Risaliti, G., Mazzarella, J.~M., Howell, J.~H., 
\& Lord, S.\ 2009, \apj, 694, 718
\bibitem[Yaqoob(1997)]{1997ApJ...479..184Y} Yaqoob, T.\ 1997, \apj, 479, 
184 
\end{thebibliography}
\end{document}